\documentclass[11pt]{article}

\usepackage{graphicx}%
\usepackage{multirow}%
\usepackage{amsmath,amssymb,amsfonts}%
\usepackage{amsthm}%
\usepackage{mathrsfs}%
\usepackage[title]{appendix}%
\usepackage{xcolor}%
\usepackage{textcomp}%
\usepackage{manyfoot}%
\usepackage{booktabs}%
\usepackage{algorithm}%
\usepackage{algorithmicx}%
\usepackage{algpseudocode}%
\usepackage{listings}%

\begin{document}

\title{Noise Models Impacts and Mitigation Strategies in Photonic Quantum Machine Learning}


\author{
A.M.A.S.D. Alagiyawanna \\
Department of Computational Mathematics \\
University of Moratuwa, Sri Lanka \\
\texttt{alagiyawannaamasd.21@uom.lk}
\and
Asoka Karunananda \\
Department of Computational Mathematics \\
University of Moratuwa, Sri Lanka \\
\texttt{asokakaru@uom.lk}
}

\date{}



\maketitle

\begin{center}
\textit{Preprint. This manuscript is currently under review at Discover Quantum Science (Springer Nature).}
\end{center}

\begin{abstract}
Photonic Quantum Machine Learning (PQML) is an emerging method to implement scalable, energy-efficient quantum information processing by combining photonic quantum computing technologies with machine learning techniques. The features of photonic technologies offer several benefits: room-temperature operation; fast (low delay) processing of signals; and the possibility of representing computations in high-dimensional (Hilbert) spaces. This makes photonic technologies a good candidate for the near-term development of quantum devices. However, noise is still a major limiting factor for the performance, reliability, and scalability of PQML implementations. This review provides a detailed and systematic analysis of the sources of noise that will affect PQML implementations. We will present an overview of the principal photonic quantum computer designs and summarize the many different types of quantum machine learning algorithms that have been successfully implemented using photonic quantum computer architectures such as variational quantum circuits, quantum neural networks, and quantum support vector machines. We identify and categorize the primary sources of noise within photonic quantum systems and how these sources of noise behave algorithm-specifically with respect to degrading the accuracy of learning, unstable training, and slower convergence than expected. Additionally, we review traditional and advanced techniques for characterizing noise and provide an extensive survey of strategies for mitigating the effects of noise on learning performance. Finally, we discuss recent advances that demonstrate PQML's capability to operate in real-world settings with realistic noise conditions and future obstacles that will challenge the use of PQML as an effective quantum processing platform.
\end{abstract}

\noindent
\textbf{arXiv categories:} quant-ph, cs.LG


\section{Introduction}\label{sec1}

Artificial Intelligence (AI) has emerged as a dominant part of modern life with many applications in diverse fields. In particular, machine learning (ML) represents a part of AI where we attempt to create algorithms for computers to learn from and subsequently predict or otherwise make decisions based on data \cite{sarker2021machine}. However, classical ML techniques typically have limitations in some combination of explainability, data insufficiency, and computational efficiency. Quantum Computing (QC) has emerged as a viable solution to the challenges posed \cite{alagiyawanna2024enhancing}.

Quantum computing utilizes the principles of quantum mechanics such as superposition and entanglement to solve problems that the classical computer cannot. 
A single qubit can be represented as a superposition of basis states:
\begin{equation}
|\psi\rangle = \alpha |0\rangle + \beta |1\rangle,
\end{equation}
where $\alpha, \beta \in \mathbb{C}$ and $|\alpha|^2 + |\beta|^2 = 1$.

Quantum states can also be represented using density matrices, which are particularly useful for describing noisy systems:
\begin{equation}
\rho = |\psi\rangle \langle \psi|
\end{equation}

For mixed states, the density matrix is given by:
\begin{equation}
\rho = \sum_i p_i |\psi_i\rangle \langle \psi_i|,
\end{equation}
where $p_i$ are probabilities such that $\sum_i p_i = 1$.

The first-ever "universal quantum computer" idea was published in 1985 by David Deutsch \cite{deutsch1985quantum}. In 1994, Peter Shor presents an algorithm that can efficiently find the factors of large numbers, which is later known as Shor's algorithm \cite{shor1994algorithms}. Quantum computation leverages the linear structure of Hilbert space, where quantum states evolve according to unitary transformations, enabling the encoding and manipulation of information in exponentially large state spaces. Entanglement further enables non-classical correlations between quantum systems, which play a crucial role in enhancing computational capabilities beyond classical limits \cite{horodecki2009quantum}.  Quantum computers can revolutionize machine learning by introducing algorithms that run much faster and can process the data more efficiently.

\begin{figure}[htbp]
\centerline{\includegraphics[width=1\linewidth]{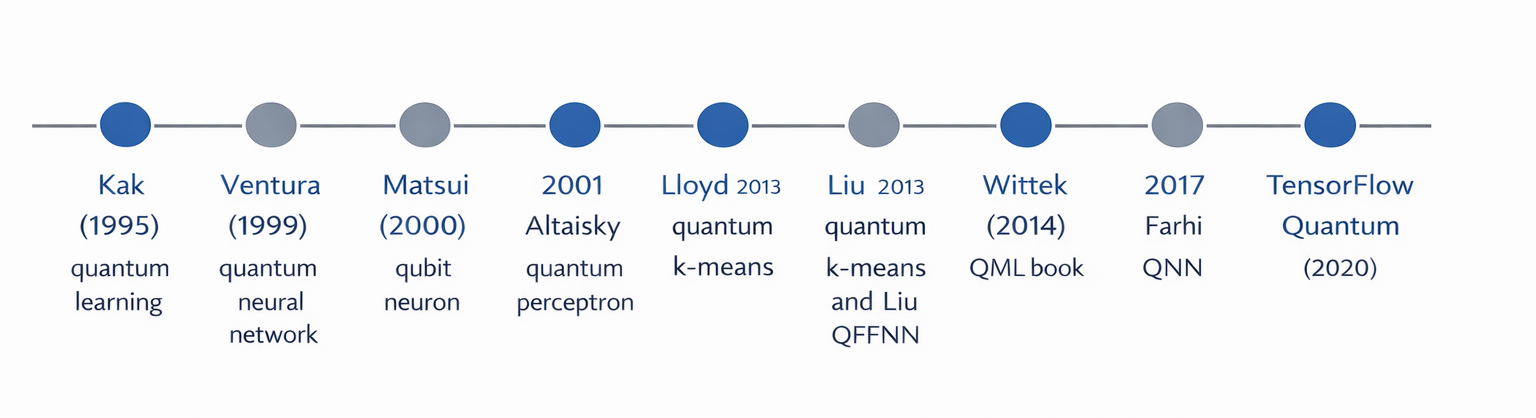}}
\caption{Timeline of major advancements in quantum machine learning.}
\label{fig1}
\end{figure}

Quantum Machine Learning (QML) is a discipline that brings together quantum computer science and traditional machine learning. There is no exact date marking the beginning of QML. But Subhash C. Kak published a paper with an idea about a biologically inspired neural network that works on quantum machine learning in 1995 \cite{kak1995quantum}. ``Figure.~\ref{fig1}'' shows the timeline of the major advancements in quantum machine learning. QML seeks to harness the special nature of quantum mechanics to make machine learning more efficient and effective. Quantum computers may have the potential to outperform classical machine learning models for certain tasks by exploiting quantum parallelism and other quantum phenomena. QML currently possesses limitations including noise in quantum systems, limited qubit counts, and the requirement for specialized hardware. However, there is ongoing research into these challenges, and QML has the potential to make great advancements to the machine learning field \cite{gil2024opportunities}.

\begin{figure}[htbp]
\centerline{\includegraphics[width=1\linewidth]{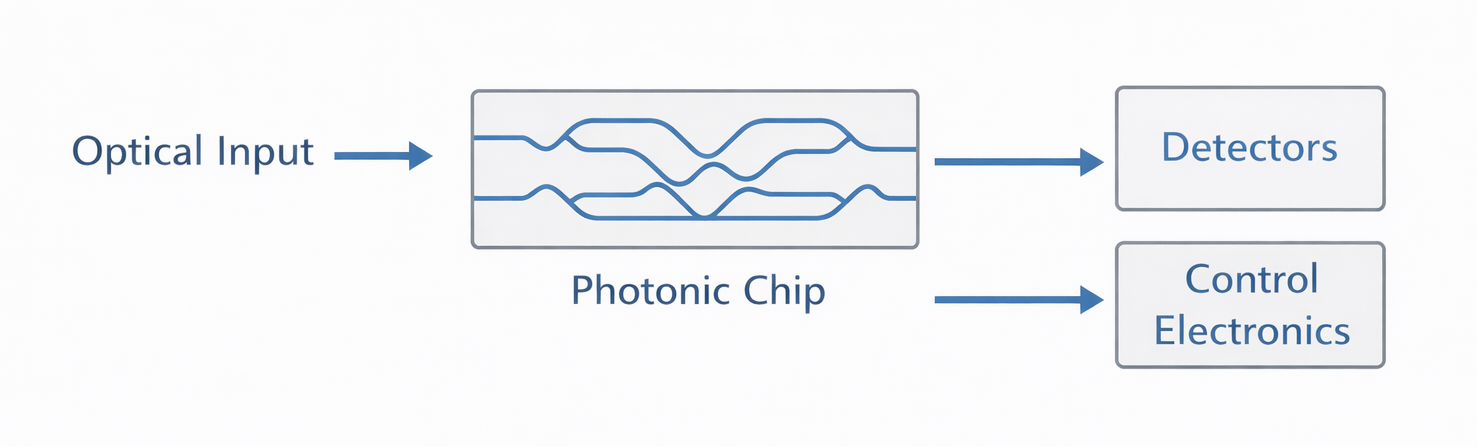}}
\caption{Schematic overview of a photonic quantum computing system showing optical input processing through a photonic chip and measurement via detectors with classical control electronics.}
\label{fig2}
\end{figure}

Among the various quantum computing platforms, photonic quantum computing has attracted particular interest due to the possibility for operation at room temperature \cite{zhong2020quantum} and because photonic systems allow low-latency real-world quantum information processing. In photonic quantum computing, qubits are denoted by photons, which are the basic units of light that can be processed by optical components like beam splitters, phase shifters, mirrors, and others to perform quantum operations. Photonic Quantum Machine Learning (PQML) combines photonic systems with quantum learning algorithms and provides an attractive methodology for potentially scalable, energy-efficient, low-resource, high-fidelity quantum computation \cite{abughanem2024photonic}. ``Figure.~\ref{fig2}'' illustrates the main components of a photonic quantum computing system.

In this review, we explore the impact of noise in photonic quantum machine learning systems. The upcoming sections of the paper are organized as follows. Section 2 discusses the fundamentals of photonic quantum machine learning. Section 3 identifies and categorizes the various noise sources in photonic quantum systems. Section 4 analyses the impact of noise on quantum machine learning algorithms. Section 5 reviews noise characterization methods. Section 6 explores noise mitigation strategies. Section 7 highlights recent advances and experimental demonstrations. Section 8 provides a discussion, where theoretical frameworks and noise models are considered in the context of recent experimental demonstrations, along with near-term applications and long-term prospects. Finally, Section 9 concludes the review.

\section{Fundamentals of Photonic Quantum Machine Learning}

Photonic quantum machine learning (PQML) is a new field that brings together photonic quantum computing with machine learning methods. In these kinds of systems, information is encoded into the quantum states of photons and manipulated by optical components to perform quantum computation \cite{blais2020quantum}. In this section, we discuss the main photonic quantum computing architectures and quantum machine learning algorithms on photonic platforms.

\subsection{Photonic Quantum Computing Architectures}

There are several architectures for photonic quantum computing, each with distinct noise characteristics and mitigation requirements. But there are only three main architectures that are widely used in photonic quantum computing. These are  discrete-variable (DV) quantum computing, continuous-variable (CV) quantum computing, and hybrid quantum computing.

\begin{figure}[htbp]
\centerline{\includegraphics[width=0.7\linewidth]{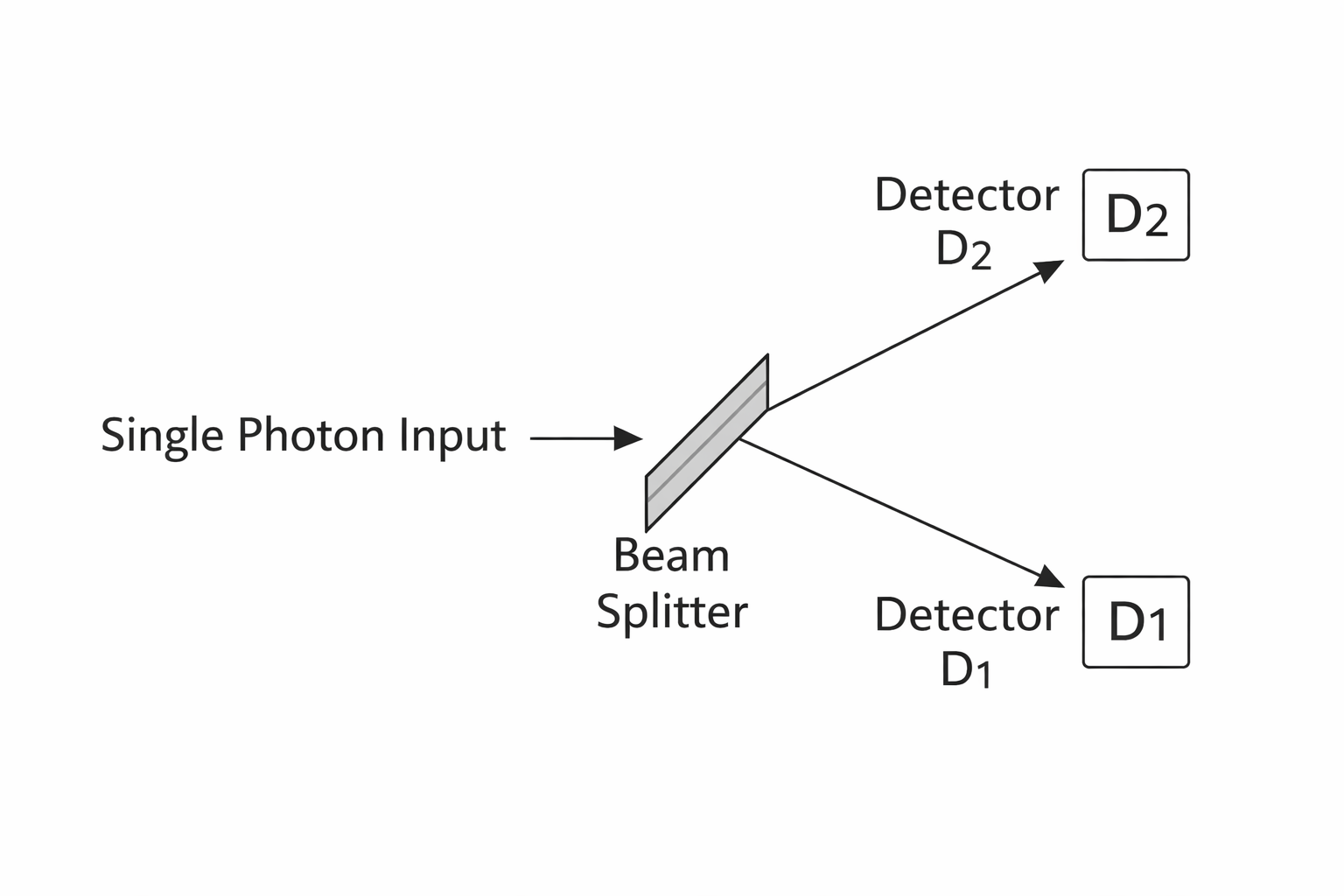}}
\caption{Schematic illustration of a single photon interacting with a beam splitter and being detected at two output ports.}
\label{fig3}
\end{figure}

Discrete Variable (DV) quantum computing involves quantum information encoded in discrete states like polarization or path, using individual photons to represent qubits \cite{lee2025photonic}. DV systems are usually implemented using single-photon sources, beam splitters (``Figure.~\ref{fig3}''), phase shifters, and single-photon detectors. In DV systems, there are many sources of noise, such as:
\begin{itemize}
    \item Photon loss: Photons can be lost due to absorption, scattering, or imperfect coupling between optical components \cite{slussarenko2019photonic}
    \item Detector inefficiencies: Single-photon detectors may have limited efficiency, leading to missed detection events \cite{kecceci2025accuracy}
    \item Mode mismatch: Imperfect alignment of optical modes can lead to reduced interference visibility and increased error rates \cite{thiele2024cryogenic}
\end{itemize}

Continuous Variable (CV) Quantum Computing encodes quantum information in continuous degrees of freedom of the infinite-dimensional Hilbert space \cite{lee2025photonic}. Rather than using discrete qubits, CV quantum computers utilize properties of light modes, such as amplitude and phase, to encode and process information. By leveraging Gaussian operations alongside non-Gaussian resources, CV quantum computing allows for universal quantum computation exploiting optical systems\cite{bourassa2021blueprint}.
In continuous-variable quantum systems, quantum information is encoded in quadrature operators:
\begin{equation}
\hat{x}, \hat{p}
\end{equation}

These operators satisfy the canonical commutation relation:
\begin{equation}
[\hat{x}, \hat{p}] = i\hbar.
\end{equation}

Noise sources in CV systems include:
\begin{itemize}
    \item Thermal noise: Interaction with the environment can introduce thermal noise, leading to decoherence and loss of quantum information \cite{simbierowicz2024inherent}
    \item Phase noise: Fluctuations in the phase of light modes can degrade the fidelity of quantum \cite{kim2022consequences}
    \item Vacuum fluctuations: The inherent quantum fluctuations in the vacuum state can introduce noise in CV systems \cite{choi2024photonic}
\end{itemize}

\begin{figure}[htbp]
\centerline{\includegraphics[width=0.85\linewidth]{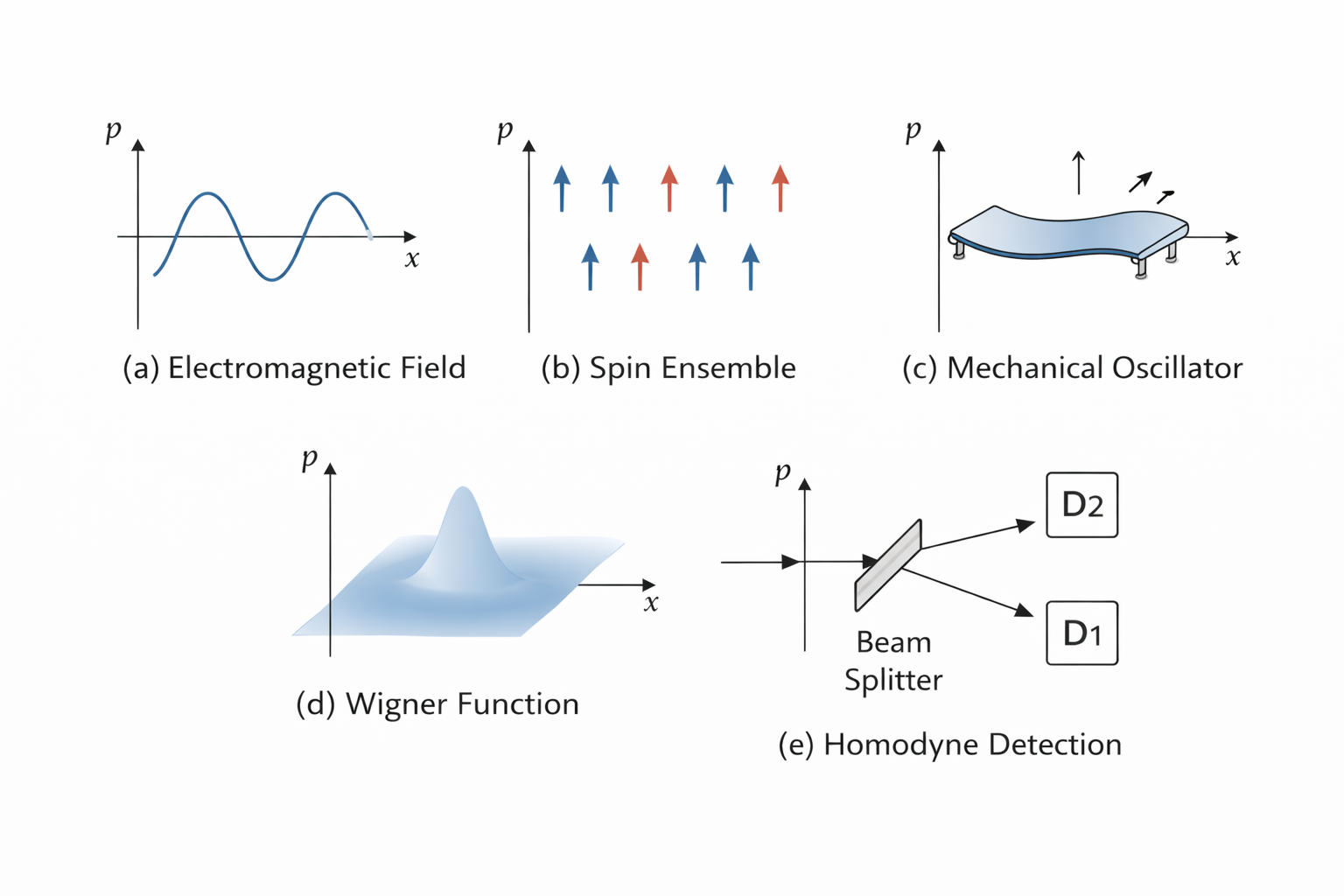}}
\caption{Schematic illustration of key components in hybrid quantum systems, including continuous-variable elements such as electromagnetic fields and mechanical oscillators, discrete-variable spin ensembles, phase-space representation via the Wigner function, and homodyne detection.}
\label{fig4}
\end{figure}

Hybrid quantum computing amalgamates the two approaches of DV and CV to capitalize on the benefits of both. In hybrid systems we can use discrete qubits encoded in single photons and CV variables for some parts of the measurement or processing. Noise sources in hybrid systems will include DV and CV noise sources, as well as additional complexities produced by the interaction of the discrete and continuous variables  \cite{zhong2020quantum}. ``Figure.~\ref{fig4}'' illustrates the building blocks of a hybrid quantum system.

\subsection{Quantum Machine Learning Algorithms on Photonic Platforms}

Various quantum machine learning algorithms have been adapted to be used on photonic platforms. Some of the prominent algorithms, such as Variational Quantum Circuits (VQC), Quantum Neural Networks (QNN), and Quantum Support Vector Machines (QSVM), have all been successfully implemented in photonic systems \cite{mohammadisavadkoohi2025systematic}.So, let's discuss these algorithms briefly.

\begin{figure}[htbp]
\centerline{\includegraphics[width=1\linewidth]{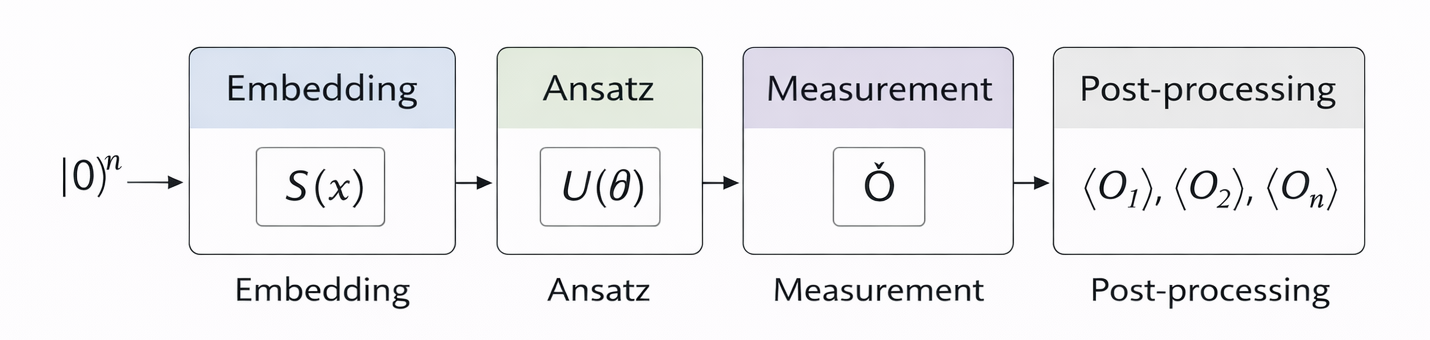}}
\caption{Schematic representation of a variational quantum circuit showing data embedding, parametrized quantum evolution, measurement, and classical post-processing.}
\label{fig5}
\end{figure}

Variational Quantum Circuits (VQC) are a powerful framework for quantum machine learning that combines classical optimization techniques and the training of quantum circuits. ``Figure.~\ref{fig5}'' illustrates a basic structure of a VQC. A photonic implementation of VQC can use the huge range of optical components to create tunable quantum circuits, which further allows VQC to successfully train and optimize quantum models \cite{sung2020using}. These variational models are particularly sensitive to noise, as the optimization process relies on accurate estimation of expectation values.

A variational quantum circuit is typically represented by a parameterized unitary:
\begin{equation}
U(\boldsymbol{\theta}) = \prod_k e^{-i \theta_k H_k},
\end{equation}
where $\boldsymbol{\theta}$ are trainable parameters and $H_k$ are Hermitian operators.

The objective of training is to minimize a cost function defined as:
\begin{equation}
C(\boldsymbol{\theta}) = \langle \psi | U^\dagger(\boldsymbol{\theta}) O U(\boldsymbol{\theta}) | \psi \rangle,
\end{equation}
where $O$ is an observable.

Quantum Neural Networks (QNN) comprise another category of algorithm that can successfully be implemented on photonic platforms. QNNs use quantum circuits for computations similar to classical neural networks, thus operating in the quantum information domain \cite{kwak2021quantum}. A key advantage of photonic QNNs is the inherent parallelism of photonic states, as well as their high-dimensionality, which benefits learning.

A quantum neural network can be expressed as a parameterized quantum transformation:
\begin{equation}
y = f\big(U(\boldsymbol{\theta}) |x\rangle\big),
\end{equation}
where $U(\boldsymbol{\theta})$ is a parameterized quantum circuit and $f$ represents measurement and classical post-processing.

Quantum Support Vector Machines (QSVM) are the quantum counterparts of classical support vector machines to be incorporated to handle tasks of classification with the help of quantum states and operations. In the photonic setting, QSVMs can take advantage of the high-dimensional Hilbert space of photons when encoding data and perform kernel evaluations based on quantum interference effects \cite{akrom2024quantum}.

Quantum support vector machines rely on quantum feature maps and kernel evaluation, defined as:
\begin{equation}
K(x_i, x_j) = |\langle \phi(x_i) | \phi(x_j) \rangle|^2,
\end{equation}
where $|\phi(x)\rangle$ represents the quantum feature encoding of input data.

\section{Noise Sources in Photonic Quantum Systems}

Photonic quantum systems are vulnerable to different sources of noise that degrade any quantum machine learning algorithm that we are conducting \cite{choi2024photonic}. Recognizing and categorizing these sources of noise is paramount to developing suitable noise mitigation strategies. In this section, we discuss the major sources of noise, broken down into two subsections, such as fundamental noise mechanisms and system-specific noise characteristics. These noise processes directly influence the performance of quantum machine learning algorithms, which we analyze in the following section.

The evolution of a quantum state under noise can be described using the Kraus operator formalism:
\begin{equation}
\rho' = \sum_k E_k \rho E_k^\dagger,
\end{equation}
where $E_k$ are Kraus operators satisfying $\sum_k E_k^\dagger E_k = I$.

Common noise models can be described within this formalism. For example, the depolarizing channel is given by:
\begin{equation}
\rho \rightarrow (1-p)\rho + \frac{p}{3}(X\rho X + Y\rho Y + Z\rho Z),
\end{equation}
where $p$ is the depolarization probability.

Amplitude damping noise can be represented using Kraus operators:
\begin{equation}
E_0 =
\begin{bmatrix}
1 & 0 \\
0 & \sqrt{1-\gamma}
\end{bmatrix}, \quad
E_1 =
\begin{bmatrix}
0 & \sqrt{\gamma} \\
0 & 0
\end{bmatrix},
\end{equation}
where $\gamma$ is the damping parameter.

These mathematical noise models provide an abstract representation of physical noise processes. In photonic quantum systems, such noise manifests through practical mechanisms such as photon loss, phase noise, and detector inefficiencies.

\subsection{Fundamental Noise Mechanisms}
While photons are excellent carriers of quantum information, they are still subject to various kinds of noise during generation, transmission, and detection, which can degrade the quality of quantum states and diminish the reliability of practitioners' quantum operations \cite{horodecki2021quantum}. Below, we identify the main types of noise in photonic quantum systems and how these types of noise affect the overall performance of a system.

One of the largest sources of noise in photonic quantum systems is \textbf{photon loss} \cite{pant2019routing}. A photon may be lost due to absorption, scattering, or improper coupling between optical elements. Photon loss reduces the number of physical qubits; this can have a serious effect on the performance of quantum algorithms \cite{couteau2023applications}. Loss occurs at many stages of quantum information processing.
\begin{itemize}
    \item \textbf{Propagation Loss}: The reduction of photons as they pass through the medium due to absorption or scattering effects \cite{palaic2024advancements}. This noise source is particularly significant in long-distance quantum communication \cite{hu2021long}.
    \item \textbf{Coupling Loss}: Losses that occur when the coupling of photons into or out of the optical element produces an inefficiency, such as when connecting to fibers \cite{kim2022consequences}. This can lead to a reduction in the overall fidelity of the quantum state.
    \item \textbf{Detector Inefficiency}: Imperfections of the single-photon detectors that will result in lost detection events or lost accurate measurement of detection events \cite{kecceci2025accuracy}. This can significantly impact the performance of quantum algorithms by introducing errors in the measurement outcomes.
\end{itemize}

Photon loss can have a detrimental effect on quantum machine learning algorithms. As an example, in variational quantum circuits, photon loss can result in an erroneous adjustment of parameters while training and exhibit poor model performance \cite{choi2024photonic}. With respect to quantum neural networks, photon loss can sever the flow of information in the network, degrading learning capabilities \cite{abbas2021power}.

\subsection{System-Specific Noise Characteristics}

The other type of noise source is related to the specific design of a particular photonic quantum device, and the way that the device itself is manufactured, like architectures and components, will also affect how the specific types of noise behave. In this subsection, many examples of photonic quantum device-specific noise have been provided.

\begin{itemize}
    \item \textbf{Mode Mismatch}:  Mismatched optical modes often cause reduced visibility of interference, which lowers the values of interference\cite{thiele2024cryogenic}. Also increased error rates on quantum algorithms that rely on interference like boson sampling, quantum walks, etc.
    \item \textbf{Phase Noise}: Fluctuating phases in a given optical mode or a particular optical undefined system will degrade the quality of a quantum state produced from interactions with that optical mode \cite{kim2022consequences}. Phase noise can be a result of many sources, such as changes in temperature of the local environment or mechanical impacts. 
    \item \textbf{Thermal Noise}: Thermal noise can come from many sources when interacting with the environment, and thermal noise will lead to decoherence, which is the complete loss of information \cite{simbierowicz2024inherent}. In continuous wave quantum key distribution systems thermal noise can degrade the mixing properties of light and also the degree to which the two or more light modes squeeze and entangle with each other.
\end{itemize}

As quantum machine learning models rely heavily on the presence of high-visibility interference, mode mismatch is one of the most detrimental sources of noise in photonic implementations. The next section will provide a thorough discussion of how mode mismatch and other types of noise represent one of the major sources of impact on photonic implementations from a practical standpoint.

\section{Impact of Noise on Quantum Machine Learning Algorithms}

There are two categories of the impact of noise. Algorithm-specific noise sensitivity deals with the impact of different types of noise on different types of quantum machine learning algorithms. Noise-resilient algorithm design focuses on creating an algorithm that will help reduce the effect of noise when using a photonic quantum system.

The expectation value of an observable is given by:
\begin{equation}
\langle O \rangle = \mathrm{Tr}(\rho O),
\end{equation}
where $\mathrm{Tr}$ denotes the trace operation. Noise modifies the quantum state $\rho$, thereby affecting the measured expectation values and degrading learning performance.

\subsection{Algorithm-Specific Noise Sensitivity}

Different quantum machine learning algorithms have varying levels of sensitivity to different types of noise. In this subsection, we analyze the impact of specific noise sources on prominent quantum machine learning algorithms implemented on photonic platforms.

Noise can have a significant impact on the operation of VQCs and QNNs. Due to their reliance on the accurate measurement of expectation values and calculation of gradients, variational quantum circuits (VQCs) are especially vulnerable to the presence of noise \cite{dangwal2023varsaw}. The impact of noise on VQC can be affected in different ways, such as:

\begin{itemize}
    \item \textbf{Gradient Vanishing/Exploding}: Noise can lead to "vanishing" or "exploding" gradients during the optimization phase, making it hard to develop a good quantum model because the gradient is either close to zero or is extremely large. \cite{mcclean2018barren}
    \item \textbf{Parameter Shift Errors}: Result of noise which causes incorrect readings, which leads to incorrect updates to the circuit parameters \cite{schuld2019evaluating}
    \item \textbf{Measurement Errors}: Errors occurring from noise in the measurement process will yield incorrect expected values; therefore, this incorrect data will not provide guidance for optimizing the quantum circuit \cite{temme2016error}
\end{itemize} 

QNNs operate using quantum mechanics and can be easily impacted by noise during their operation. Therefore, noise may impact the ability of QNNs to learn and adapt. Similarly, feedforward structures, convolutional, and recurrent QNNs will each experience issues with noise at various levels. So we have to use different strategies to mitigate these noise effects.

\subsection{Noise-Resilient Algorithm Design}

Noise-resilient algorithm design focuses on creating quantum machine learning algorithms that can withstand the presence of noise in photonic quantum systems. According to recent research, we can get mainly two types: hardware-aware compilation and algorithmic error suppression. ``Figure.~\ref{fig6}'' illustrates the concept of quantum error mitigation (QEM), which aims to design noise-resilient algorithms.

Hardware-aware compiling tools for modern quantum circuits take advantage of the specific types of noise present in the underlying hardware to create quantum circuits that are as close to optimal as possible. Specifically, the optimized circuit designs reduce the number of overall errors based on the type of noise present in the photonic platform used. The optimizations include reducing circuit depths and limiting the number of entangling operations within a particular circuit design to reduce the negative effect of photon loss and decoherence.

Through a variety of methods, algorithmic error suppression techniques are utilized to mitigate the negative effects of noise on QML. These include error mitigation methods, for example, probabilistic error cancellation, zero-noise extrapolation, and symmetry verification. In addition, photonic quantum systems have used these methods to increase the reliability of QML by allowing for greater resilience to the presence of noise.

\section{Noise Characterization Methods}

Characterizing noise is an important part of understanding and reducing the impact of noise in quantum photonic systems. Accurately characterizing the source of noise enables the creation of efficient methods that reduce noise and improve the efficiency of quantum machine learning algorithms. In this section, we will present both the traditional methods used to characterize noise and advanced techniques that can be applied to analyze noise in quantum photonic systems.

\subsection{Traditional Noise Characterization Approaches}

These approaches are widely used to identify and quantify noise in photonic quantum systems. Some of the most common traditional methods include:

\begin{itemize}
    \item \textbf{Quantum Process Tomography (QPT)}: A method used to fully characterize quantum operations of photonic quantum systems. The idea is to prepare some amount of previously prepared states, perform an operation on these states with quantum operations, and then measure the output states after the operation to obtain a reconstruction of the process matrix\cite{chuang1997prescription}.
    \item \textbf{Randomized Benchmarking (RB)}: A technique that evaluates the average error rates for quantum gates by applying random gate sequences and measuring the final state fidelity. This is used to assess the performance of gates under noise conditions \cite{magesan2011scalable}.
    \item \textbf{Noise Spectroscopy}: A method that analyzes the overview of the noise frequency spectra in quantum systems by applying certain pulse sequences and measuring the response of a quantum system to those pulse sequences. The resulting noise using these techniques provides information about the dominant sources of noise and their frequency characteristics \cite{bylander2011noise}. ``Figure.~\ref{fig7}'' shows a comparison of noise spectra reconstructed.
\end{itemize}

Due to the fact that certain traditional approaches will not capture the complete spectrum of complexities associated with noise arising from interactions in photonic quantum systems, there is a need for advanced characterization techniques designed to enhance understanding of the underlying causes of these types of noise.

\subsection{Advanced Characterisation Techniques}

To improve upon the limitations of traditional approaches, researchers have begun employing advanced techniques for characterizing noise in photonic quantum systems. The following are some of the advanced techniques used for characterizing noise:

\begin{itemize}
    \item \textbf{Machine Learning-Based Noise Characterization}: These approaches can analyze large amounts of data collected from quantum experiments and discover patterns and relationships within the data associated with the fluctuations of noise. Therefore, machine learning methods provide a way to model complex noise phenomena that cannot be modeled using traditional methods\cite{torlai2018neural}.
    \item \textbf{Adaptive Tomography}:  In this form of quantum process tomography, measurement strategies are used to maximize the amount of information gathered about the noise process through adaptive sequencing. Thus, this technique provides a more efficient method for characterizing noise with fewer measurements than conventional techniques \cite{glos2022adaptive}.
    \item \textbf{Bayesian Inference Methods}: These techniques allow for estimating unknown noise parameters based on previous knowledge and updating estimates as experimental data becomes available. Consequently, Bayesian methods yield a probabilistic understanding of the characteristics of noise sources \cite{lange2023adaptive}.
\end{itemize}

After characterizing the noise using the above techniques, we need to consider developing effective noise mitigation strategies to reduce the impact of noise on photonic quantum machine learning systems.

\section{Noise Mitigation Strategies}

A major goal when improving the performance of photonic quantum machine learning devices is to mitigate noise. If noise can be successfully managed, it will decrease the adverse effects that noise has on the reliability of quantum algorithms. This section will cover different types of noise-mitigation strategies that may be applicable to photonic quantum devices.

\subsection{Hardware-Level Mitigation}

Mitigation strategies at the hardware level focus on improving the quality of the physical components that comprise the photonic quantum system as a method for eliminating noise. One of the most effective methods to reduce noise is by incorporating quality optical components like optical waveguides with high-efficiency single-photon detectors and stable laser sources with low losses. By implementing low-loss waveguides, high-efficiency single-photon detectors, and stable laser sources, it becomes possible to effectively reduce the number of photons detected due to inefficiencies at the detector and lost to the environment due to extinction of the light \cite{slussarenko2019photonic}. Likewise, the incorporation of active stability techniques like feedback control systems will further enhance the stability of optical components, thereby eliminating phase noise \cite{kim2022consequences}.

\subsection{Encoding-Based Mitigation}

Encoding-based mitigation techniques are focused on creating a set of specific types of encoded agencies to help protect quantum information from degrading environmental disturbance or noise. One encoding method for quantum information is called \textbf{Redundant Encoding}, which is a way of spreading quantum information throughout multiple photons or modes of light within a physical composite quantum \cite{browne2005resource}. Redundant structured encoding spreads the quantum information out, providing more flexibility with regard to the loss of photons. The second form of encoding method is referred to as the use of \textbf{Decoherence-Free Subspaces(DFS)}, which involves utilizing specific subspaces in the quantum region in order to encode the quantum information with immunity from a number of different forms of environmental noise, specifically forms of noise created through the combined dephasing of our universe \cite{lidar1998decoherence}. Another encoding technique for encoding quantum information is the implementation of \textbf{Quantum Error-Correcting Codes}, which detect and correct errors that occur through environmental noise during quantum processes. Each of the above encoding techniques works to increase the resilience of quantum machine learning systems to a combination of different environmental noise sources.

\subsection{Algorithmic Mitigation}

\begin{figure}[htbp]
\centerline{\includegraphics[width=1\linewidth]{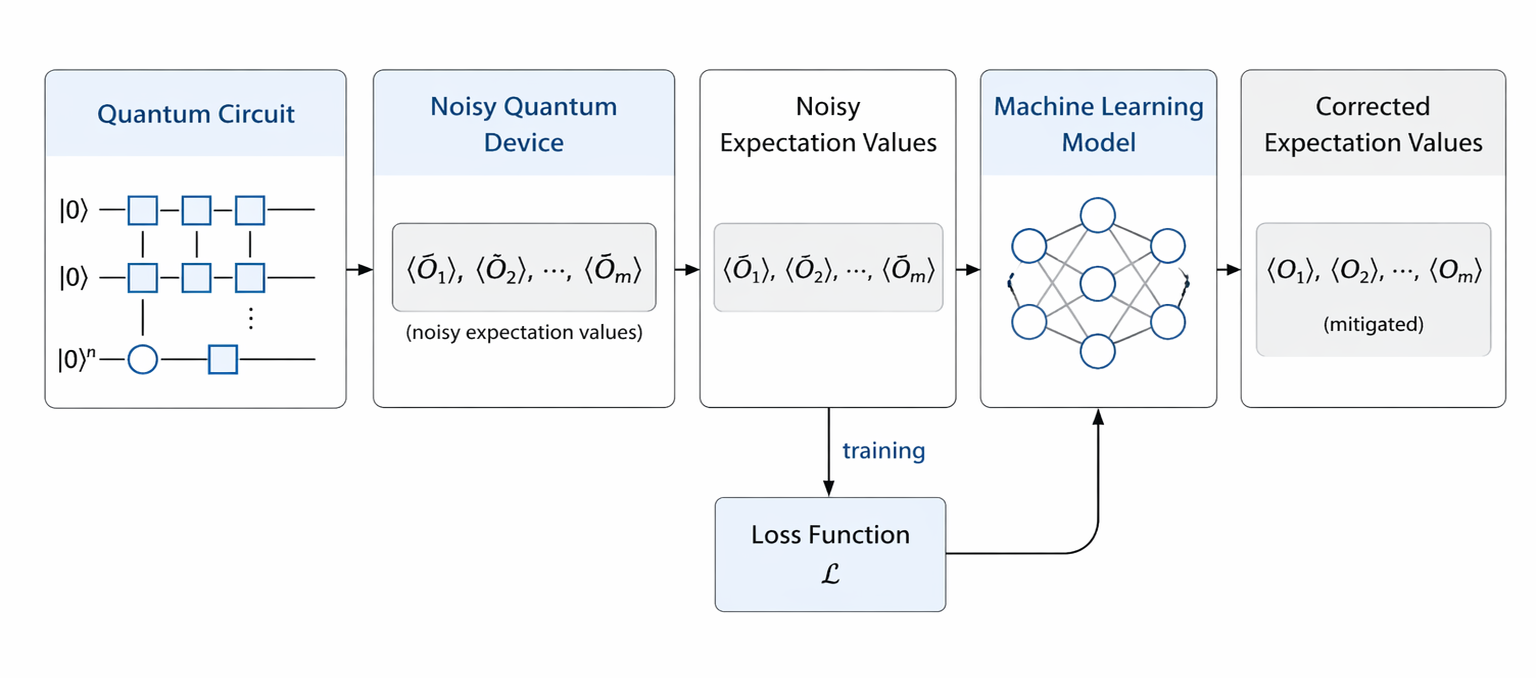}}
\caption{Schematic illustration of machine learning-based quantum error mitigation where a trained model learns to map noisy expectation values obtained from a quantum circuit to noise-mitigated predictions through supervised training.}
\label{fig8}
\end{figure}

Algorithmic mitigation techniques aim to build quantum machine learning algorithms that can stay resilient when faced with noise in photonic quantum systems. To that end, \textbf{error mitigation methods} are available; these are designed to help eliminate the effects of noise on the output of quantum algorithms and therefore do not have to perform complete error correction. Error mitigation methods include probabilistic error cancellation, zero-noise extrapolation, and symmetry verification \cite{temme2016error}. Another approach used by algorithmic mitigation is to create \textbf{noise-resilient algorithms} that have been created specifically to handle certain noise types well. Noise-aware training is a technique that allows a user to train an algorithm to understand what type of noise is present during its learning phase. By combining these algorithmic mitigation techniques, we are able to improve quantum machine learning algorithm performance despite the presence of noise. Other than above approaches, \textbf{Machine Learning-Based Noise Reduction} techniques can also be used to reduce noise in photonic quantum systems. These techniques use machine learning models to identify and filter out noise from quantum data, thereby improving the quality of the quantum information being processed \cite{torlai2018neural}. ``Figure.~\ref{fig8}'' illustrates such a machine-learning-based quantum error mitigation approach.

\subsection{Hybrid Classical-Quantum Approaches}

Hybrid classical-quantum approaches are strategies that merge classical and quantum computing technologies to reduce this interference. A hybrid classical-quantum approach can also be achieved through \textbf{Classical Post-Processing}, which uses classical algorithms for analysis and correction of the output data from quantum algorithms that have been degraded by noise as a result of their execution in a noisy environment \cite{temme2016error}. On the other hand, \textbf{variational hybrid algorithms} can also be hybrid algorithms based on a combination of quantum computing and classical optimization techniques to develop adaptive algorithms that can continually adjust to the noise throughout their execution \cite{sung2020using}. Hybrid classical-quantum approaches use a combination of the advantages of both classical and quantum computing to develop effective ways to reduce the effect of noise in the development of photonic quantum machine learning systems.

The combination of the above strategies will reduce noise on photonic quantum machine learning systems through more efficient operations and more consistent operation of the entire system.

\section{Recent Advances and Experimental Demonstrations}

Recent years have seen significant advancements in photonic quantum machine learning, with numerous experimental demonstrations showcasing the potential of photonic systems for quantum information processing. 

2016 was a seminal year for continuous-variable (CV) photonic quantum computing, as it was the year in which a \textbf{1 million-mode photonic cluster state} was demonstrated, thus demonstrating the scale-up capabilities of CV photonic systems \cite{yoshikawa2016invited}. Using this very large-scale cluster state, a variety of quantum information processing tasks were also demonstrated (e.g., quantum teleportation and entanglement distribution).

\begin{figure}[htbp]
\centerline{\includegraphics[width=1\linewidth]{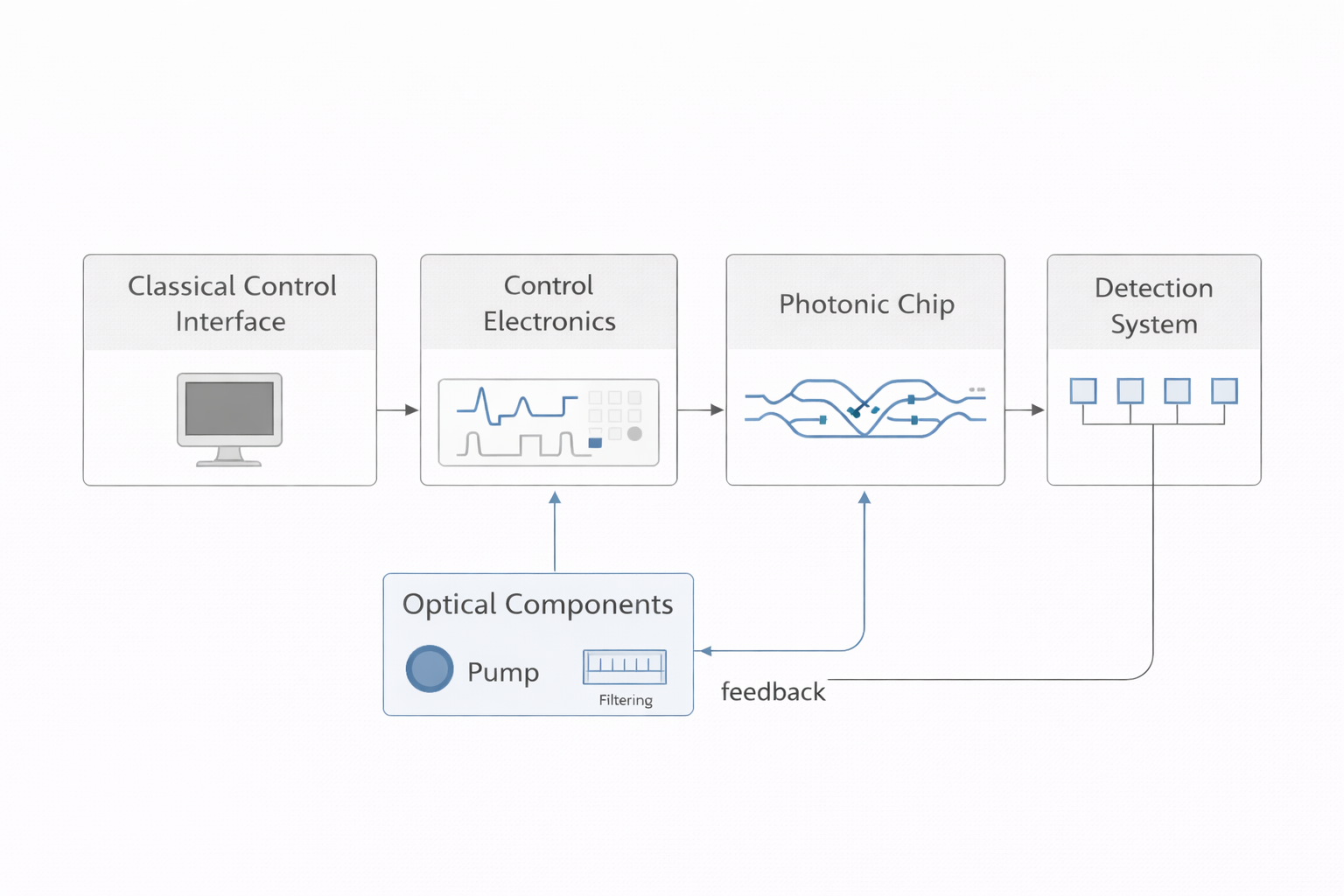}}
\caption{System-level schematic of an integrated programmable photonic quantum processor, showing the interaction between a classical control interface, control electronics, photonic chip, optical components, and detection system, with feedback for system optimization.}
\label{fig9}
\end{figure}

Starting in 2021, research teams have developed \textbf{integrated programmable photonic processors} at multiple institutions (Xanadu, NIST, and MIT), thus enabling multiple quantum algorithms to be implemented on a single photonic chip, which enables reconfigurable and scalable photonic quantum computing \cite{arrazola2021quantum}. In addition, the integrated programmable photonic processors also reduced system noise by decreasing both the number of optical components and improving the stability of the overall system. ``Figure.~\ref{fig9}'' illustrates an overview of a processor.

Experimentation of a \textbf{photonic quantum neural network} was demonstrated in 2023, using a programmable photonic chip to perform classifications over quantum data, indicating that an innovation is possible in using photonics in quantum ML applications \cite{pai2023power}. Advantages of photonic systems include very low latencies and working within room temperature conditions.

To continue developments toward 2024, a new \textbf{superconducting-photonics hybrid chip} was conceived, allowing for ~99\% loss resilience at photon levels and leading to the reduction of the leading type of noise in photonic systems \cite{lee2024fault}.  These developments will allow for a much larger scale of practical deployment of photonic-based quantum computing.

These recent advances and experimental demonstrations highlight the progress made in photonic quantum machine learning, showcasing the potential of photonic systems for scalable, low-noise quantum information processing.

\section{Discussion}

According to the above sections, we can discuss three main points, such as theoretical frameworks and noise modeling perspectives, near-term applications of PQML under noise, and long-term prospects and open challenges. These models are commonly expressed using quantum channels such as the Kraus representation introduced earlier.

\subsection{Theoretical Frameworks and Noise Modeling Perspectives}

To develop the most promising strategies to reduce or eliminate noise in photonic quantum systems, we must construct a theoretical foundation that accurately captures the types of noise found in these systems. Noise may have many different effects on the state of a quantum system; thus, there are several types of noise models that have been proposed to define how it impacts a quantum state, such as amplitude and phase damping and depolarizing channels \cite{choi2024photonic}. By accurately modeling the various noise processes affecting photonic quantum systems, we may be able to predict how each noise type will act upon different types of quantum algorithms, as well as design robust quantum algorithms that are resilient to all types of noise. Furthermore, advanced techniques for modeling noise, including the development of stochastic noise models and non-Markovian noise models \cite{bylander2011noise}, provide the most accurate representation of the many types of noise that exist in the real world and will allow us to study the temporal correlations in noise, the memory effects associated with it, and the long-term behavior of quantum states in noisy environments.

\subsection{Near-Term Applications of PQML under Noise}

Advances in photonic quantum machine learning (PQML) will bring about a transformational impact in many areas, including but not limited to areas that are susceptible to noise. For example, due to its nature of being able to classify chemicals, photos, videos, etc. in real time, PQML could allow for more precise image classification than classical classification techniques \cite{mohammadisavadkoohi2025systematic}. A good example of this would be using a PQML algorithm for classifying chemicals found in samples of murky water, which cannot be done effectively by current methods of classification. As PQML continues to evolve, researchers will find even greater opportunities to utilize the incredible capabilities of PQML to their advantage, particularly due to the inherent robustness of PQML digital circuitry and room temperature operation.

\subsection{Long-Term Prospects and Open Challenges}

The future of photonic quantum machine learning (PQML) looks bright, but many current challenges are preventing this field from reaching its maximum potential. Among these challenges are the limitations on the scalability of quantum photon systems. Although considerable work has been completed in building integrated photonic circuits, creating systems that may undertake more qubits or continue functioning at a low noise level when built larger will be difficult to develop \cite{arrazola2021quantum}. Additionally, integrating photonic quantum systems with other methods for executing quantum technologies such as superconducting qubit systems or an ion confined to a magnetic field presents not only opportunities to use the advantages of each unique quantum scheme but also creates additional issues associated with noise control and system coherence \cite{andersen2015hybrid}. Therefore, addressing these challenges will require increased interdisciplinary cooperation among physicists, engineers, and computer scientists to advance the state-of-the-art of PQML further.

\section{Conclusion}

Our paper provided a comprehensive review of the state of research into quantum machine learning (QML) and photonic quantum computing (PQC) that focused on the noise limitation in the quantum machine learning systems. We have covered the general principles related to photonic quantum machine learning, including the various types of photonic quantum computing architectures, algorithms developed using quantum machine learning on photonic platforms, and various noise sources in photonic quantum systems and their classifications, including natural noise mechanisms and the characteristics associated with the photonic quantum systems. Also, we provided an overview of how the various forms of noise affect the quantum machine learning algorithm/algorithms and discussed the current research into algorithm-specific noise sensitivity and the design of noise-resilient algorithms.

Additionally, this review has included an overview of the various types of methods and current status of research into the various methods and current research into noise characterization, as well as how the various techniques, including hardware-level techniques to develop encoding-based, algorithmic, and hybrid classical-quantum methods, and experimental demonstrations work. Overall, this article highlights not only the challenges and achievements in managing the problems caused by noise within photonic quantum machine learning systems, but it also provides an extensive overview of the state of research into noise in photonic quantum machine learning and identifies significant obstacles and future avenues of research for realizing feasible, scalable, and effective photonic quantum machine learning applications on the photonic quantum machine learning platforms.

\section*{List of Abbreviations}

AI — Artificial Intelligence  

CV — Continuous Variable  

DV — Discrete Variable  

ML — Machine Learning  

QC — Quantum Computing  

QEM — Quantum Error Mitigation  

QML — Quantum Machine Learning  

QNN — Quantum Neural Network  

PQML — Photonic Quantum Machine Learning  

QSVM — Quantum Support Vector Machine  

VQC — Variational Quantum Circuit


\bibliographystyle{unsrt}
\bibliography{sn-bibliography}

\end{document}